# EQUI JOIN QUERY ACCELERATION USING ALGEBRAIC SIGNATURES (published in IADIS'2008)

Riad Mokadem, Abdelkader Hameurlain, Franck Morvan
*Institut de Recherche en Informatique de Toulouse (IRIT), Université Paul Sabatier*
*118, route de Narbonne, 31062 Toulouse Cedex, France*

**ABSTRACT**

Evaluation of join queries is very challenging since they have to deal with an increasing data size. We study the relational join query processing realized by hash tables and we focus on the case of equi join queries. We propose to use a new form of signatures, the algebraic signatures, for fast comparison between values of two attributes in relations participating in an equi join operations. Our technique is efficient especially when the attribute join is a long string. In this paper, we investigate this issue and prove that algebraic signatures combined to known hash join technique constitute an efficient method to accelerate equi join operations. Algebraic signatures allow fast string search. They are descending from the Karp-Rabin signatures. String matching using our algebraic calculus is then several times faster comparing to the fastest known methods, e.g. Boyer Moore.
We justify our approach and present an experimental evaluation. We also present a cost analysis for an equi join operation using algebraic signatures. The performance evaluation of our technique shows the improvement of query processing times. We also discuss the reductions of required memory sizes and the disk I/O. The main contribution of this paper is the using of algebraic signatures to accelerate equi join operations especially when the attribute join is a long string and to avoid multiples I/O disk by reduce memory requirement.

**KEYWORDS**

Hash Join Algorithms, Algebraic Signatures, String Matching, Performance Evaluation.

## 1. INTRODUCTION

Join is one of the most expensive but fundamental query operation. Hash join queries have been the study of many query optimization efforts over the past two decades. Compared to other join strategies, a hash-based join algorithm is particularly efficient and easily parallelized for join computation. It is commonly used in today's commercial database systems to implement equi joins efficiently. However, multiple disk I/O, memory requirement and skew effect of data can severely deteriorate the system performance. Several works treated this problem and studies join processing techniques.

For this purpose, several hash join algorithms were proposed: simple hash join, grace hash join and hybrid hash join but it's not always possible during optimization to determine which hash join is used. [2] compared the sort merge, simple hash, grace hash and hybrid hash algorithms and concluded that hybrid hash has the lowest cost. Other works [12] presented a new data structure for general join query acceleration, the hit-list, and an algorithm for its use. The hit-list is a surrogate index providing the mapping between the values of two attributes in a relation participating in an equijoin. Many recent studies [13] have also focused on improving the cache performance and indexing techniques in query acceleration. Disadvantages of this approach include the maintenance costs, navigation costs as the indexes grow large, paging costs when forming the join, and the need to enforce strict domain integrity constraints across the whole database. Also, caches improve latency only if requested data is found in cache. More recently, [1] presented new prefetching techniques to increase speedups for the join and partition phases over grace hash join algorithm. It shows that hash join algorithms suffer from extensive CPU cache stalls. It shows also that the Grace hash join algorithm spends over 73% of its user time stalled on CPU cache misses, and explores the use of prefetching to improve its cache performance.

On the other hand, database systems provide powerful query languages for structured data, but are not particularly well suited for querying text information and especially for string matching operations used in equi join queries.

For this purpose, we propose in this article a novel algorithm based on algebraic signatures and combined with a known hash join algorithm. The proposed algorithm exploits the values of join attributes in tuples and hashed them by simple values called signatures. One compares the signature of the searched string with that of the currently examined string join attribute. In the following sections, only the equi join is discussed. Experiments show that our algorithm reduce the cost of performing a join by using algebraic signatures in order to accelerate comparison notably to compare large strings join attributes. Algebraic signatures are very fast and constitute an important advance in domain of string matching [6]. Search using algebraic signatures was up to eleven times faster for ASCII documents, seventy times for DNA documents and about six times for XML documents.

Our method presents two advantages. First, it allows fast comparisons. The experiments show it is typically several times faster than simple hash join algorithm in particular when comparing long strings. Then, using algebraic signatures permit to reduce the required memory size. In consequence, it also reduces I/O transfers. We consider this paper to be our first step in acceleration of join operations.

This paper is organized as follows: in section 2, we present hash join algorithms and focus in simple and grace hash join algorithms. We introduce problems related to complex string matching operations in equi join queries and also problem of memory requirement for the existing algorithms. Then, we present the algebraic signatures and discuss using of them for fast string matching and contribution to accelerate equi join operations. Section 4 present analytic cost model for both grace hash join and our proposed hash join algorithm based in algebraic signatures. We also compare our equi join execution times using algebraic signatures to the simple hash join times. We conclude with direction for the further work especially for using the algebraic signatures in other join techniques, e.g. semi join based join algorithm.

## 2. HASH JOIN ALGORITHMS AND PROBLEM POSITION

Basically, the join operation merges two relations. Let R and S be two relations participating in the join operation J. Let $m=|R|$ and $n=|S|$ sizes of relations R and S respectively. Hash join algorithm requires an equi joins predicate to join tables via two attributes r and s (respectively from R and S). Hash joins are better when there is a significant difference in the sizes of the tables being joined. Let size of R be smaller than that of S ($m<n$). To execute join operation between R and S, simple hash Join is executed on two phases: Build and Probe.

The algorithm first builds a hash table on the smaller relation. An in-memory hash table H is build. It's based on the value of applying a hash function (h) to the join attributes of R. The hash table performs like an index and consists of an array of hash buckets, each composed of a header and (possibly) an array of hash cells pointed by the header. It's the build phase. Then, in the probe phase, each tuple of the other larger relation are used to probe the hash table by means of applying the same hash function h to its joint attribute. Tuples from each relation that match on join attribute are concatenated and written to result relation. When the main memory available to a hash join is too small to hold the build relation and the hash table, the simplistic algorithm suffers from excessive random disk accesses. Various other schemes have been proposed. The Grace algorithm introduced by Kitsuregawa differs from the simple hash algorithm in its data partitioning phase of the two joining relations. It introduces dynamic decisions about bucket partitioning and then creates temporary relations on disks. Instead of reading individual tuples, several blocks (partitions) are read from an input relation before switching to the other to avoid excessive random I/Os. These partitions have important characteristic that all tuples with the same join attribute value will share the same bucket. Pairs of build and probe partitions are then joined separately as in the simple algorithm and build the result relation (join phase). Another algorithm, hybrid hash join [2, 9] is combination of the Grace hash and the simple hash algorithms. The primary difference is that the hybrid hash join algorithm reserves an area of memory to join records in during the partitioning phase.

However, hash joins are CPU-intensive in comparison to nested loops, computationally expensive and difficult to optimize [1]. They are especially affected by available memory and applying prefetching to hash joins is complicated by the data dependencies, multiple code paths, and inherent randomness of hashing. The existing solutions are almost too expensive. A join operation also faces the performance degradation by data skew. Since the source relations usually result from earlier operations such as selection or projection, data distribution is hard to predict. So buckets tend to vary in size and then, bucket overflow occurs. This requires extra I/O to repartition buckets to be smaller than the memory. In consequence, each tuple may be frequently read and written. Performance is diminished. In order to treat these problems, we present a new hash join algorithm based on algebraic signatures. We show that using algebraic signatures decrease memory size need to join the two relations and then, permits less I/O disks. The needed main memory will be decrease since we represent each tuple by signature of its join attribute. More memory pages are available for other blocks and we don't have to access them by I/O disk operations.

Also, instead to compare attributes values character per character, we only compare their signatures. We explore this for join operations and focus on the equi join queries. Join attribute is a long string. For ease of presentation, we describe two examples using an equi join operation. Let us consider query taken from benchmark TPC-D for the Product Type Profit Measure: "Finds, for each nation and each year, the profit for all Parts ordered in that year which contain a specified substring in their part-names and which were filled by a Supplier in that nation". This query is a 6-table join with selection criteria applied to only one table. A text search is done against some of these tables and string matching is applied in equi joins operations. Observe also that the number of matching invoked depends on the cardinality of the relational tables. Therefore, for huge relational tables, the tuple substitution method will be prohibitively expensive. Cache can be used to remember the set of all past probes but for huge relations, it will be insufficient. So, string matching cost can be reduced by hashing values of comparing attributes. This is why we introduce algebraic signatures. Let us view another example. Let us consider two simple relations to join: relation *Student (N_Student, Name_University)* and relation *University (Name_University, town)*. Lets we interest now to students studying in Toulouse. Join attribute here is the Name_University attribute which can be a long string (e.g, 'Paul Sabatier University'). The basic method is the sequential match over the successive symbols of the visited attribute. Another scheme consists to encode or compress this attribute. In our algorithm, we introduce the using of algebraic signatures for fast string matching. Our scan performs without decoding any data. We hash each join attribute and then, compare only the values of its algebraic signatures. In following section, we show that matching in the probe phase is accelerate when we hash each join attribute by its algebraic signatures and then, comparing between these values.

## 3. ALGEBRAIC SIGNATURES FOR A HASH JOIN ALGORITHMS

We introduce algebraic signatures to accelerate comparison in equi join operations executing in probe phase of hash join algorithm. These signatures are also generated in the build step when hash table is constructing. Let show a brief view of algebraic signatures. A signature is a string of a few bytes intended to identify uniquely the contents of a data object (tuple, page, file, etc.). Different signatures prove the inequality of the contents, while same signatures indicate their equality with overwhelmingly high probability.

## 3.1 Algebraic Signatures

An algebraic signature consists of n symbols, where a symbol is either a byte or a double byte, though theoretically symbols can be bit strings of any length f. We call it algebraic signatures because it uses Galois Field calculations. Galois field (GF) is finite field. We denote $GF(2^f)$ a GF over the set of all binary strings of a certain length f. Every non-zero element is power of a primitive element α. In GF calculation, the sum of two strings is implemented by *XOR* operation. To implement the multiplication in GF, we implement logarithms and antilogarithms tables described in [7]. One major characteristic, new for any known signature scheme, is certain detection of small changes of parameterized size. More precisely, we detect for sure any change that does not exceed n-symbols for an n-symbol signature. For larger changes, the collision probability is typically negligible, as for the other known schemes. We sign our objects with 4-byte signatures, instead of 20-bytes standard SHA-1 signatures that would be impractical for us. We call page P a string of $l$ symbols $p_i$ ; $i = 0...l-1$. In our case, symbols $p_i$ are byte (2-bytes words resp). The symbols are elements of a Galois field, GF ($2^f$) (f =8, 16 resp). We assume that $l < 2^f -1$. Let α= ($α^1, α^2,.., α^n$ ) be a vector of different non-zero elements of the Galois field. We call the n-symbol signature base, or simply the base.

The (n-symbol) P signature or, simply, P signature, based on α, is the vector
$sig_\alpha(P) = (sig_\alpha^1(P), sig_\alpha^2(P),..., sig_\alpha^n(P))$   Where for each α, we set $sig_\alpha(P) = \Sigma p_i \alpha^i$  $i=1...n$
Our n-symbol algebraic signature is the concatenation of n power series of GF symbols in $GF(2^{16})$. While our algebraic signatures are not cryptographically secure as is SHA1, they exhibit a number of attractive properties. First, we may produce short signatures, sufficient for our goals, e.g., only 4B long. Next, the scheme is the first known, to the best of our knowledge, to guarantee that objects differing by n symbols have guaranteed different n-symbol signatures. Furthermore, the probability that a switch leads to a collision is also sufficiently low, namely $2^{-nf}$ with signature length f in bits.

## 3.2 Hash Join Algorithm Using Algebraic Signatures

Equi joins acceleration results from two factors of using algebraic signatures: the speed of calculus of these signatures and the restriction of I/O because of the size of our signatures. So, hash join algorithms don't need to have more buckets in hard disk. This is allowed by full the elimination of paging and the reduction of access hard disk in order to search other tuples participating in the final result especially when we have huge relations.  Algebraic signatures are very efficient for string matching. [6] compares the speed of string matching algorithm using algebraic signatures to the Boyer-Moore scheme, usually the fastest method known. Search using algebraic signatures was up to eleven times faster for ASCII documents. That is why we concentrate in the equi join operation dealing with strings. Accordingly, algebraic signatures are very useful in case of equi join when join attribute is a long string. Since the signature calculation for equi join is built on top of hash join algorithm, we have presented the simple hash join algorithm first and then extend it to the using of algebraic signatures. We analyze the dependencies in a hash table visit in the join phase to show why a naive comparison of tuples would slow join operation. In our approach, we'll interest in previous computation of signatures during the build phase and comparison of signatures during the probe phase.

We motivate our use of signatures in this context as follows. Our signatures are calculated in build phase. This calculation is doing just once while tuples are not update since algebraic signatures detect any change in tuples. Each signature of join attribute in each tuple of R is calculate and put into the hash table according to the value obtained after applying the hash function to this join attribute. After having built the hash-table, for each tuple of S, the same hash function is also applied to the join attribute. If the hash value indicates an empty entry in the hash-table, this tuple and its signature are ignored. Otherwise, the signature of join attribute in S is calculated. Matching of join attribute signatures is very fast since signatures are only 4Bytes where string they signed could be dozen bytes.

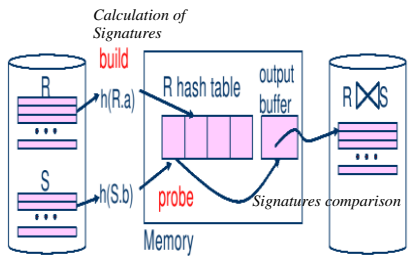

Let R, S be two relations to join. T be the join's result
Sign j(r), Sign j(s) be algebraic signatures values of join attribute j calculate on tuples r, s in R and S respectively.
$H_{sign}$ be hash table containing signatures values of R.

    For (each tuple r in R)  Pre calculate Sign j(r);
    Build (R, $H_{sign}$);
    For (each tuple s in S) {
      Probe ($H_{Sign}$, S);
      If (any signature match is found)    Output the matching tuples;
          }
    Materialize (T);

**Figure 1**: Signatures matching      **Figure 2**:  Hash join algorithm using algebraic signatures

## 4.  PERFORMANCE EVALUATION

In this section we are interested in the performance evaluation of our approach. We evaluate the performance of our algorithms in order to evaluate the robustness of the analytical study. In the following section, the cost formulas estimating the memory requirement, I/O disk and response time of our hash join using algebraic signatures will be detailed. In the rest of this section, we describe the experimental environment and then analyze the obtained results. We present our results in three forms: (a) we report the performance results of our algebraic signatures and advantage of using them in equi join operations. (b) we also report the result of our algorithm in case of variation of string join attribute size and (c) we also experiments the advantage of using algebraic signatures to reduce memory size necessary to perform hash join operation.

## 4.1 Cost Analysis for Hash Join Algorithm Using Algebraic Signatures

Efficient join algorithm cost models are important, as they are needed by relational query optimizers (which employ cost-based optimization algorithms) in order to derive efficient processing plans for relational queries. As a starting point, accurate analytical models that capture centralized query performance are required. Our focus here is the development of a model for hash join algorithm based on algebraic signatures. We believe that a meaningful cost model can be built upon only three characteristic parameter sets, describing main memory size, the I/O bandwidth and CPU execution of join operation. For each parameter, we discuss the advantage of using algebraic signatures to improve cost of executing join operation. To illustrate the profit of using algebraic signatures in executing of join operations, we illustrate a relatively analytic model and then compare it to previous cost model of hash join algorithms. Simple analytic cost models for hash based join algorithms were first presented in [9]. These models only count the total number of page I/0 operations in the algorithm. In the world of Hash Joins, there are many different methods to cope with insufficient memory. Workload of site in local processing depends on: available memory, used memory, number of I/O per second, active processes number. We consider here all these parameters.

*Memory_Needed= Max(b/k, k) pages .... (0)*
*# Disk I/Os = 2 b + 4(b-r) .... (1)*
*Build_buckets = ( m\* l_f + n\* l_f )\* (Disk_Input_R+ hash_bucket+ Disk Output buckets) .... (2)*
*Build_hash= m\*(Read_bucket+ **Signature generating**+**Build_Sign_cost**)+ Write_$H_{sign}$_to_disk .... (3)*
*Probe_hash=n\*(Disk Input S + **Signature Calculation** + **Probe_Sign_cost**) .... (4)*
*Output_result= Disk Input tuples result + Materialize T .... (5)*
*Alg_Sign_Hash_Join_Cost = Memory_Needed + #Disk I/Os + Build_buckets + Build_hash + Probe_hash+ Output_result .... (6)*

**Figure 3**: Total hash join cost using algebraic signatures

The methods we would like to introduce here in order to compare it to our algorithm is the Grace Hash Join algorithms. Let R, S be base relations, containing n and m tuples each, stored in *b* blocks on a disk in area of size *r* to build the hash table for the first bucket. Suppose that m<<n. Because R and S are written once into a temporary file, we need *6b* disk accesses (*read R, S, store $R_{temp}$, $S_{temp}$, read $R_{temp}$, $S_{temp}$*) [4]. The minimum amount of memory for this algorithm is show in formula (0) [4]. If we assume optimal memory utilization, the first bucket has the maximal size of *r = k - b* and the other buckets are also of maximal size k. In this configuration, the number of disk I/Os is the number of reads for the participating tables plus I/Os for writing and reading the buckets *2 ... b+1* to a temporary file [4]. This cost is given by formula (1). Formula (2) shows the cost to build different buckets when R not fit in memory. *Disk_Input_R* consists to read tuple from disk. Each bucket is hashed in main memory (using up to *k* blocks) and one input buffer for the outer relation is used. This cost is given by formula (2). In the third phase, buckets of relation R are read and an in-memory hash table $H_{Sign}$ is filled with the tuples of the buckets. The other terms used in Figure 3 describing total cost for hash join algorithm using algebraic signatures have the following meaning:

*build_Sign_cost*: The cost of building the hash table $H_{Sign}$ described in our proposed algorithm (Figure3).
*probe_Sign_cost*: The cost to probe $H_{Sign}$. We only compare algebraic signatures values of join attributes.
*l_f*     : Loop factor which is number of loops necessary to partitioning relations and build hash buckets.
Formulas estimating the memory requirement, response time and I/O disks costs of a join of 2 relations T = Join (R,S) by using the hash join algorithm based on algebraic signatures are in figure 3.
In build phase, algebraic signature is calculated for attribute join in each tuple of R (step 3). When we compare our build and probe costs (*Build_Sign_cost* and *Probe_Sign_cost*) to the simple build and probe costs described by [9] (*Build_cost* and *Probe_cost*), our build cost is slightly higher since we calculate algebraic signature for each join attribute in each tuple. However, algebraic signatures are very fast to generate (20-30 *ms* per 1 MB date) [7]. Also, the saving is much important in the probe phase since we compare only algebraic signatures values instead of comparing attributes character per character. Then, hash table $H_{Sign}$ is build. When this table is greater than memory, it is partitioned and parts of $H_{Sign}$ are written to disk. In formula (4), the 'Probe_Sign_Cost' phase corresponds to signatures comparison. The hash table is probed for every tuple of S (formula 4). We discuss this step later in this section. If the tuples match, a result is created. Otherwise, it's ignored. In probe step, the probe is doing by comparison of algebraic signatures. The cost described by formula (5) consists to load the tuples result and materialize T. The formula (6) shows the total hash join cost based on algebraic signatures. In our algorithm using algebraic signatures, it's not necessary to read complete tuples. Signatures are calculated for only join attribute of each tuple. The needed main memory will be decrease since we represent each tuple by signature of its join attribute. Because of small size of these signatures, T is smallest and cost described by (0) is reduced. More pages are available for other blocks and we don't have to access them by I/O disk operations. Comparing to traditional join hash algorithm, in hash join algorithm based algebraic signatures we have an additional cost in formula (3) to compute algebraic signatures for attribute join of each tuple in R. Only signatures of join attribute is store in memory. Then, more number of join attribute tuples is available in main memory rather than in a disk file and therefore less I/O to disk are necessary. Then, cost described by formula (1) is also reduced. In formula (4), we calculate algebraic signature of each tuple in S and then, probe the hash table build in step described by formula (3) by comparing algebraic signatures values. Using algebraic signatures permit fast comparison. Instead of comparing dozen of bytes like in traditional scheme, we compare only values of algebraic signatures to probe S. Experiments show that signatures comparison is fast.

## 4.2 Experimental Environment

Let us consider two relations R and S residing in same site, with number of rows estimated at 10000 and 20000 respectively (Table R has smaller size). T= Join (R, S) result of hash simple join. The selectivity factor of join here is 1.5/max [3, 10]. The average number of tuples per page (4KB) is estimated to 32 for 128B row. We conducted the experiments on a single CPU computer (Linux Redhat, processor speed 2.8 Ghz, cache=256 KB, memory 512 MB). The tuples inserted in the two relations are randomly generated (data are integer, character, string and date) where the join string attribute is a string. The hash join based on algebraic signatures is implemented in java and compiled using the GNU javac compiler. The experiments are evaluated on algorithm described in section 2. We compare our results to the standard hash join result. For this purpose, we experiment the variation of the join attribute size and thereafter, the size of the two relations.

## 4.3 Performance Analysis and Comparisons

Three primary factors affects relational operations: computation costs in CPU, disk access costs and memory size available. It is important to know the effect of CPU times compared with the I/O disk times. Commonly, the disk access costs are dominant. Accelerate treatment is also important. We estimate the CPU costs by number of tuples to be handled and the disk access costs by calculating the number disk pages to be accessed. In our experiments, elapsed times for the two joins algorithms include the time required to write the final relation to disk. We have compared our results to times obtained by applying the hash join algorithm. In Figure 4, we show the estimated performance of both hash join algorithm and the one using algebraic signatures for same experiments. We have varied size of our join attribute. This attribute can be a string like the example we have noted in section 2. The results show that, at join with factor of selectivity 1.5, our algorithm is better when the attribute join exceed 7 Bytes. For small string, 2Bytes for example, the cost of comparing strings character per character is less than to generate and then compare algebraic signatures. The gain increases for larger join attribute string as it should. Our algorithm is appears particularly efficient for longer attribute join strings in the context of performing equi joins. Figure 4 shows that the curve corresponding to the simple hash join algorithm grew linear with the join attribute size while the curve corresponding to our algorithm using algebraic signatures is growing less because of the fix size of our signatures (4 Bytes in our case). Figure 5 presents speedup according to join execution. The speedup here is the ratio of execution join time using algebraic signatures over the execution time using simple hash join algorithm. The using of algebraic signatures can advantageous for join string attribute de jointure exceeding 7Bytes. This ratio is about 1.2 for 10B attribute join string and linearly increases to 2 and 2.8 for 50B and 100B attribute join string respectively. This is directly related to gain of signature calculus opposed to compare each join attributes characters in the two relations. The traditional comparison is implemented by the *a.equal(b)* java command. For huge relations R and S (eg. 100000 tuples), this ratio is much higher. This is due to several accesses to disk since we have several buckets for each relation. Also, hash table T does not fit in memory in case of using simple hash join algorithm. In consequence, many I/O are generating. When we experiments with |R|= 100000 tuples and S=|200000|, the ratio between elapsed time increase into 4. One of reason is due for reason that hash table T not fit in memory.

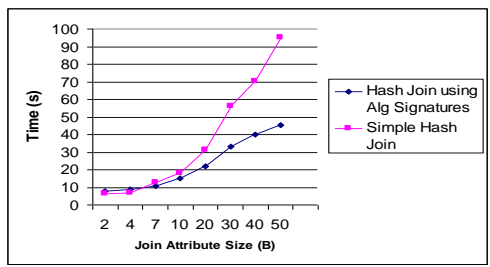
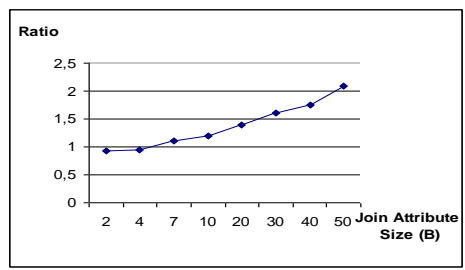

**Figure 4**: Join execution times          **Figure 5**: Speed up (HJ over HJ using alg sign)

Also, we have observed that improvement is more important when we consider only comparison times without read the two relations and without materialization of join relation result (pipeline join operation). The reason is that read R and S from disk and write the join result T to disk constitute dominant times in total join cost. We have observed this result since we reduced join selectivity factor and then reduced the output of result to disk. However, we consider this result to be preliminary as we have not yet compared performance of our algorithm with the other pipeline techniques for equi join operations. Memory size consumption is affected by using algebraic signatures in executing equi join operation. Figure 6 shows the values of memory requirement obtained by TOP program used as measurement tool for measuring the MEM parameter. The performance of simple hash join algorithm is significantly affected by the amount of available memory and performs well when the smaller relation is less than twice the size of available memory. In our experiments, bucket overflow did not occur in the latest tests of both algorithms. We measure the average of available free memory for each experiment. Figure 7 shows the average memory used for execution of both hash algorithms (with/ without using algebraic signatures). In this experiment, we vary number of tuples in relations R (|S|=20000 tuples). Introduction of our algebraic signatures allows more main memory since only algebraic signatures of join attributes suffice to compare tuples. Note that Figure 6 and figure 7 don't take in account the I/O operations necessary to read the input relations R and S. Also in case of voluminous relations R and S, we have a gain in the memory size when we used our algorithm based on algebraic signatures. It permit to reduce the number of I/Os by avoid access to the several buckets of each relation. We have also less I/O to and from disk. Also, for a given page size, the calculation times for $sig_{\alpha,n}$ were linear in n. Depending of number of tuples, the gain will be high while comparisons concerns very long strings attributes. The actual calculation times of the 4B long signature $sig_{\alpha,2}$ (calculated in $GF(2^{16})$) was 20-30 *ms* per 1 MB of RAM data. The calculus signature time grew linear with the tuple size while signatures comparison time for varied strings is unchanged because of the fix size of our signatures.

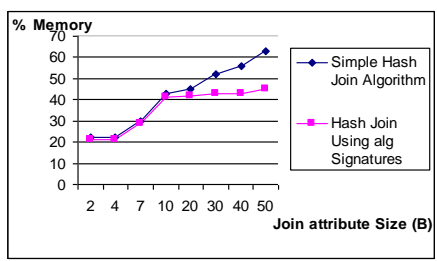
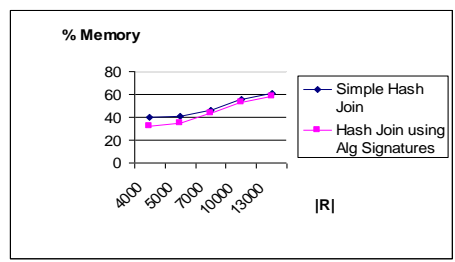

**Figure 6**: Memory requirement for join attribute variation          **Figure 7**: Memory requirement for greater relations

## 5. CONCLUSION AND FUTURE WORKS

In this paper, we have proposed a novel algorithm based on algebraic signatures and combined with a known hash join algorithm. It can be used in other kinds of join operations like the left and right joins which are frequent database operations. Algebraic signatures are very efficient to compare strings. We studied its impact to accelerate join query operations. Our method shows the potential for fast equi joins queries in databases. In the performance evaluation, our experiments show that hash join algorithm based on algebraic signatures can perform very well when comparing to other hash join algorithms especially for long string join attribute.

In distributed environments, the execution cost of a join operation is constituted by computation CPU, I/O and communication costs. It is important to consider all these costs to achieve an optimal performance. In this paper, our method shows that algebraic signatures perform the computation cost and reduce I/O disk operations. In consequence, it reduces the memory requirements. Combined to several join techniques, algebraic signatures can reduce communication resources often the dominant factor in a distributed environment. Finally, the outcome fully of the experimental study confirms our theoretical proposal. Our algorithm provides good performance to avoid limitations of memory size and reduce I/O transfers. The speed of string matches of the algebraic signatures seems to open an interesting perspective for the join database operations.

An extension of the work reported in this study is to compare our results to the Grace hash join ones. We have presented its cost model in section 4. The preliminary conclusion when we analyze the formulas presented in same section is that our algebraic signatures have good impact and can significantly reduce the I/O cost. We also plan to consider other hash join techniques. We are also exploring methods to increase the speed of the signature calculation. Cumulative signatures constitute another method for using our signatures [7]. Preliminary results suggest that it is two to three times faster than the method reported here. Another extension of the work reported in this study consist to experiment the benefit of using algebraic signatures for the left, right and bushy joins which are frequent database operations. Also, using algebraic signatures in semi join based join algorithm can also improve our equi join query acceleration. It can reduce the amount of data to transmit between sites. Since, it keeps the network traffic low because of reducing size of algebraic signatures.